# On Determining the Spectrum of Primordial Inhomogeneity from the $COBE^1$ DMR Sky Maps: II. Results of Two Year Data Analysis

K.M. Górski[2,3,4], G. Hinshaw[5], A.J. Banday[2], C.L. Bennett[6],
E.L. Wright[7], A. Kogut[5], G.F. Smoot[8] & P. Lubin[9]

[1]The National Aeronautics and Space Administration/Goddard Space Flight Center (NASA/GSFC) is responsible for the design, development, and operation of the *Cosmic Background Explorer (COBE)*. Scientific guidance is provided by the *COBE* Science Working Group. GSFC is also responsible for the development of the analysis software and for the production of the mission data sets.

[2]Universities Space Research Association, NASA/Goddard Space Flight Center, Code 685, Greenbelt, Maryland 20771.

[3]on leave from Warsaw University Observatory, Aleje Ujazdowskie 4, 00-478 Warszawa, Poland.

[4]e-mail: *gorski@stars.gsfc.nasa.gov*

[5]Hughes STX Corporation, 4400 Forbes Blvd., Lanham MD 20706.

[6]NASA Goddard Space Flight Center, Code 685, Greenbelt MD 20771.

[7]UCLA Astronomy Department, Los Angeles CA 90024-1562.

[8]LBL, SSL, & CfPA, Bldg 50-351, University of California, Berkeley CA 94720.

[9]UCSB Physics Department, Santa Barbara CA 93106.

## ABSTRACT

A new technique of Fourier analysis on a cut sky (Górski, 1994) has been applied to the two year $COBE$ DMR sky maps. The Bayesian power spectrum estimation results are consistent with the Harrison-Zel'dovich $n = 1$ model. The maximum likelihood estimates of the parameters of the power spectrum of primordial perturbations are $n = 1.22$ (1.02) and $Q_{rms-PS} = 17$ (20) $\mu$K including (excluding) the quadrupole. The marginal likelihood function on $n$ renders $n = 1.10 \pm 0.32$ ($0.87 \pm 0.36$).

*Subject headings:* cosmic microwave background — cosmology: observations

## 1. INTRODUCTION

The determination of the power spectrum of primordial inhomogeneities and their consistency with the predictions of inflation are critical issues in contemporary cosmology. Standard inflationary scenarios imply a power law spectrum $P(k) \propto k^n$, with $n \simeq 1$ on the scales probed in the $COBE$ DMR sky maps. Previous attempts to determine the primordial power spectrum from the DMR anisotropy data (Smoot *et al.* 1992, Wright *et al.* 1992, Adams *et al.* 1992, Scaramella & Vittorio 1993, Seljak & Bertschinger 1993, Bennett, *et al.* 1994, Smoot *et al.* 1994, Wright, *et al.* 1994a, Bond 1994) found both relatively steep spectra and sensitivity to the inclusion or exclusion of the quadrupole in their results. These methods have all employed approximate statistical techniques and have relied on Monte Carlo techniques to assess and/or calibrate the final results.

In this *Letter*, the $COBE$ DMR two year 53 and 90 GHz data are analysed to determine the primordial power spectrum using the method described in Górski (1994). Specifically, the sky maps are Fourier decomposed in the basis of orthonormal functions on the Galaxy cut sky to yield a set of harmonic mode coefficients. These are then used in a maximum likelihood analysis to infer the parameters of the theoretical anisotropy models. The merits of the present method are: 1) harmonic mode coupling is explicitly accounted for by construction of the orthonormal functions on the Galaxy-cut sky, 2) since the harmonic modes have a Gaussian probability distribution (Górski, 1994) an *exact* likelihood function for the model parameters can be employed, 3) the technique permits a simultaneous analysis of different frequency maps taking full advantage of both the auto- and cross-correlation information in the data.

In this analysis we Fourier decompose the two year DMR sky maps over the spectral range up to $\ell = 30$, where both the DMR beam response has fallen to $\sim 0.2$ (Wright *et al.* 1994b) and the multipole amplitude is noise dominated. The monopole and dipole components, which are physically irrelevant for the power spectrum estimation, are algebraically excluded (Górski 1994). In addition, no attempt is made to formally model and subtract the high latitude galactic emission, which is predominantly quadrupolar in nature. Therefore, in what follows, the power spectrum parameters are derived for two cases: one in which data spanning the multipole range $\ell \in [2, 30]$ is used, and the other in which the quadrupole ($\ell = 2$) mode is excluded.

Hereafter, bold, upper case letters denote matrices; bold, lower case letters denote vectors; and $p$ is a pixel label. Specific equations from Górski (1994) will be referred to by G immediately followed by the equation number.

## 2. DATA

Each individual DMR sky map is a collection of pixelized measurements of temperature perturbations, $\Delta(p)$, comprising cosmological, receiver noise, galactic, and systematic signals. The instrument noise is characterized by the number of observations per pixel, and an rms noise per observation, $\sigma_0$ (derived from Table 1 of Bennett *et al.* 1994). In order to limit the effects of galactic emission on the power spectrum determination, a $|b| = 20°$ Galaxy cut is imposed on the maps. Systematic effects have been strongly limited (Bennett *et al.* 1994) and are neglected in the present analysis, as are the weak high latitude galactic signals. The 53 and 90 GHz two year cut-sky maps are analyzed in this work. We form weighted average maps at each frequency by combining the individual A and B channels as follows

$$\Delta_{53} = 0.578\,\Delta_{53A} + 0.422\,\Delta_{53B},$$
$$\Delta_{90} = 0.366\,\Delta_{90A} + 0.634\,\Delta_{90B}, \qquad (1)$$

where all temperatures are in thermodynamic units. These weights were chosen to minimize the average noise variance per pixel in the combined maps. Fourier analysis of the cut-sky maps is performed in the 961-dimensional linear space, spanned by the orthonormal functions $\boldsymbol{\psi}$ constructed in Górski (1994). The harmonic coefficients of a map in the $\boldsymbol{\psi}$ basis are given by (see G8 and G3)

$$\Delta(p)|_{p \in \{cut\ sky\}} = \mathbf{c}^T \cdot \boldsymbol{\psi}(p), \quad \text{where} \quad \mathbf{c} = \langle \Delta\,\boldsymbol{\psi} \rangle_{\{cut\ sky\}}. \qquad (2)$$

The data vector for joint analysis of the 53 and 90 GHz data was formed according to $\mathbf{c}_{53\oplus90}^T = (\mathbf{c}_{53}^T, \mathbf{c}_{90}^T)$.

## 3. LIKELIHOOD ANALYSIS

The covariance matrices, $\mathbf{C} = \mathbf{C}_{CMB} + \mathbf{C}_N$, of the harmonic amplitudes, $\mathbf{c} = \mathbf{c}_{CMB} + \mathbf{c}_N$, specify the Gaussian probability distribution (G15) of the receiver noise contaminated theoretical CMB anisotropy signal. The power spectrum dependent matrix $\mathbf{C}_{CMB}$ can be conveniently parameterized by two quantities, $Q_{rms-PS}$ and $n$ (see G10, and G11).

The receiver noise is modeled as a spatially uncorrelated Gaussian random process with a frequency dependent variance per pixel $\sigma^2(p) = \langle \Delta_N^2(p) \rangle_{\{noise\ ensemble\}} = \sigma_0^2/N(p)$. Eq. G13 was used to obtain the Fourier space noise correlation matrices $\mathbf{C}_N$. The resulting rms noise amplitude per harmonic mode is $4.8\mu K$ for the 53 GHz data, and $7.8\mu K$ for the 90 GHz data. These values are important for an a priori assessment of the sensitivity of DMR to the anisotropy predicted by a particular cosmology.

Having established an appropriate mathematical representation of the data, specifically the Fourier amplitude vector $\mathbf{c}_{53\oplus 90}$, the anisotropy correlation matrix $\mathbf{C}_{CMB}$ and the noise correlation matrices $\mathbf{C}_{N53}$ and $\mathbf{C}_{N90}$, we can perform a Bayesian likelihood analysis to determine the power spectrum parameters $Q_{rms-PS}$ and $n$. The necessary ingredients for a proper Bayesian statistical analysis (e.g. Berger 1980) of values of $Q_{rms-PS}$ and $n$ to be inferred from the data are the likelihood function, $P(\mathbf{c}_{53\oplus 90}|Q_{rms-PS},n)$, and the prior distribution $\pi(Q_{rms-PS},n)$. These then define the posterior distribution

$$P(Q_{rms-PS},n|\mathbf{c}_{53\oplus 90}) \propto \pi(Q_{rms-PS},n) \times P(\mathbf{c}_{53\oplus 90}|Q_{rms-PS},n). \tag{3}$$

We evaluate the likelihood function for $Q_{rms-PS}$ and $n$ (see eqs. G15 and G16) given data either including or excluding the quadrupole anisotropy (which involves correlation matrices of dimension 1914, or 1904, respectively).

Figure 1 shows two views of the likelihood function evaluated with the quadrupole anisotropy included; the no quadrupole case yields qualitatively similar contours. This likelihood function is confined to a very narrow and steep ridge in the parameter space conveniently defined by the multipole amplitude $a_9 = 8\ \mu K$. Clearly, any prior $\pi(Q_{rms-PS},n)$ that is slowly varying near the peak of the likelihood will not significantly affect the inference on $Q_{rms-PS}$ and $n$. We have tested the sensitivity of the derived maximum likelihood parameter values and credible intervals to some specific assumptions about $\pi$. The following prior distributions were considered: (1) $\pi = const$, (2) $\pi \propto 1/(1+n)$, (3) $\pi \propto (1+n)$, (4) $\pi \propto 1/Q_{rms-PS}$, (5) $\pi \propto Q_{rms-PS}$, (6) $\pi \propto 1/(1+n)/Q_{rms-PS}$, and (7) $\pi \propto (1+n)Q_{rms-PS}$. As expected the results are robust to $\sim 1\ \mu K$ in $Q_{rms-PS}$, and $\sim 0.1$ in $n$. Hence, in the following we adopt a constant prior distribution.

Figure 2 shows the 68%, 95%, and 99.7% contours for the posterior distributions of $Q_{rms-PS}$ and $n$ evaluated with and without the quadrupole anisotropy. Exclusion of the quadrupole from the analysis results in considerable elongation of the likelihood contours towards low $n$ and high $Q_{rms-PS}$ values. This is the region of the parameter space seriously constrained by the inclusion of the relatively low observed quadrupole (Bennett *et al.* 1994). Marginalization by integration of the likelihood function over $Q_{rms-PS}$ renders a likelihood density for the spectral index $n$ shown in Fig. 3. The conditional likelihood density on $Q_{rms-PS}$ for $n = 1$ (the Harrison-Zel'dovich spectrum) is shown in Fig. 4. Relevant numbers are presented in Table 1.

## 4. DISCUSSION

Miscellaneous issues pertaining to the proper assessment of our results include (1) noise model uncertainties, (2) the question of biases in parameter inference, and (3) comparison to other attempts at determination of the power spectrum.

The method of power spectrum determination presented in this *Letter* requires an adequate description of the statistical properties of noise in the sky maps. In order to test the correctness of construction of the Gaussian noise ensemble, we have evaluated the Fourier components of the cut-sky two year difference maps, $\mathbf{c}_{A-B}$, at both frequencies, and the a priori difference map noise covariance matrices (see eq. G13), $\mathbf{C}_{N(A-B)}$. The reduced $\chi^2$, defined as $\mathbf{c}_{A-B}^T \cdot \mathbf{C}_{N(A-B)} \cdot \mathbf{c}_{A-B}/N$ where $N = 957$, is 0.93 at 53 GHz and 1.06 at 90 GHz. Therefore, the a priori noise model appears sufficiently accurate when constructed using the flight rms noise values. Nevertheless, the noise rms per observation, $\sigma_0$, is only known to an accuracy of $\sim 1\%$. The possibility of a mismatch between the assumed and the actual noise levels can perturb the spectral index and amplitude estimation. Specifically, an underestimate of the noise leaves some small scale power which the likelihood method translates into a steeper spectrum than required by the true sky anisotropy, or *vice versa*. If $\sigma_0$ is adjusted by $\sim 1\%$, the maximum likelihood values for $Q_{rms-PS}$ and $n$ are shifted by $\sim 0.5$ $\mu$K and $\sim 0.04$, respectively, insignificant amounts when compared to the overall uncertainty in the inferred spectrum.

In principle, maximum likelihood estimates of non-linear model parameters are only asymptotically unbiased. In order to test whether any biases exist in this analysis, Monte Carlo simulations of noisy random skies drawn from an ensemble of fixed $Q_{rms-PS}$ and $n$ and with the noise properties of the two year 53 GHz maps were analysed. No statistically significant bias in the simulated sample-averaged estimates of the parameters $Q_{rms-PS}$ and $n$ was detected when the quadrupole was either included or excluded. A sub-ensemble of simulations with suppressed quadrupole amplitudes was also considered: in this case, a general bias was observed to lower $Q_{rms-PS}$ and higher $n$. Given that the observed sky quadrupole is low with respect to our inferred ensemble averaged value of $Q_{rms-PS} = 20$ $\mu$K, this affords a plausible explanation for the small $\delta n \sim 0.2$ difference between spectral slopes inferred from the sky maps with or without the quadrupole (see Table 1).

By choosing $\ell_{max} = 30$ we have analysed the *COBE* DMR maps over a spectral range larger than previously implemented. Other methods are inherently insensitive to the high-$\ell$ information content of the data either by construction (e.g. Wright *et al.* 1994, where $\ell_{max} = 19$ was adopted), or as a consequence of additional binning or smoothing of the data (e.g. Bennett *et al.* 1994, Smoot *et al.* 1994). Specifically, we have tested our method over the restricted range $\ell \in [3, 19]$ as a direct comparison with Wright *et al.* (1994). The maximum likelihood estimates in this case are $Q_{rms-PS} = 17.8$ $\mu$K, and $n = 1.22$. The steepening of the resulting spectrum, in accord with Wright *et al.* (1994), is driven by the relatively high amplitude of several modes within $\ell \in [14, 19]$ in the 53 GHz sum map. In addition, we find that the power in the corresponding bin of the 53 GHz (A-B) difference map falls below the a priori noise model, thus Wright *et al.* determine a still higher power in their quadratic power spectrum estimator. With the inclusion of $\ell \geq 20$ modes the effect of this excess is more effectively constrained and attributed to noise. A more detailed comparison of the application of different methods of power spectrum inference is deferred to a future paper.

In summary, the new, statistically unbiased technique of power spectrum determination described in Górski (1994) has been implemented for the *COBE* DMR 53 and 90 GHz two year sky maps. The results are completely consistent with the Harrison-Zel'dovich $n = 1$ spectrum favored by inflationary models. If this is indeed the case, then the two year *COBE* DMR data provide a $\sim 12\sigma$ significant detection, $Q_{rms-PS} = 20~\mu K$, for the amplitude of the spectrum of primordial inhomogeneity.

We gratefully acknowledge the efforts of those contributing to the *COBE* DMR. *COBE* is supported by the Office of Space Sciences of NASA Headquarters.

Table 1: Derived parameters from the likelihood analysis

|  | including Quadrupole | | excluding Quadrupole | |
|---|---|---|---|---|
|  | $Q_{rms-PS}$ ($\mu$K) | $n$ | $Q_{rms-PS}$ ($\mu$K) | $n$ |
| Maximum likelihood values | 17.0 | 1.22 | 20.0 | 1.02 |
| 68% credible interval | [12.2, 24.6] | [0.70, 1.68] | [13.5, 30.5] | [0.43, 1.55] |
| 95% credible interval | [10.0, 31.8] | [0.35, 1.95] | [10.6, 41.5] | [0.02, 1.85] |
| Marginal likelihood on $n$ | 1.10 $\pm$ 0.32 | | 0.87 $\pm$ 0.36 | |
| Conditional likelihood on $Q_{rms-PS}^{n=1}$ | 19.9 $\pm$ 1.6 | | 20.4 $\pm$ 1.7 | |

---

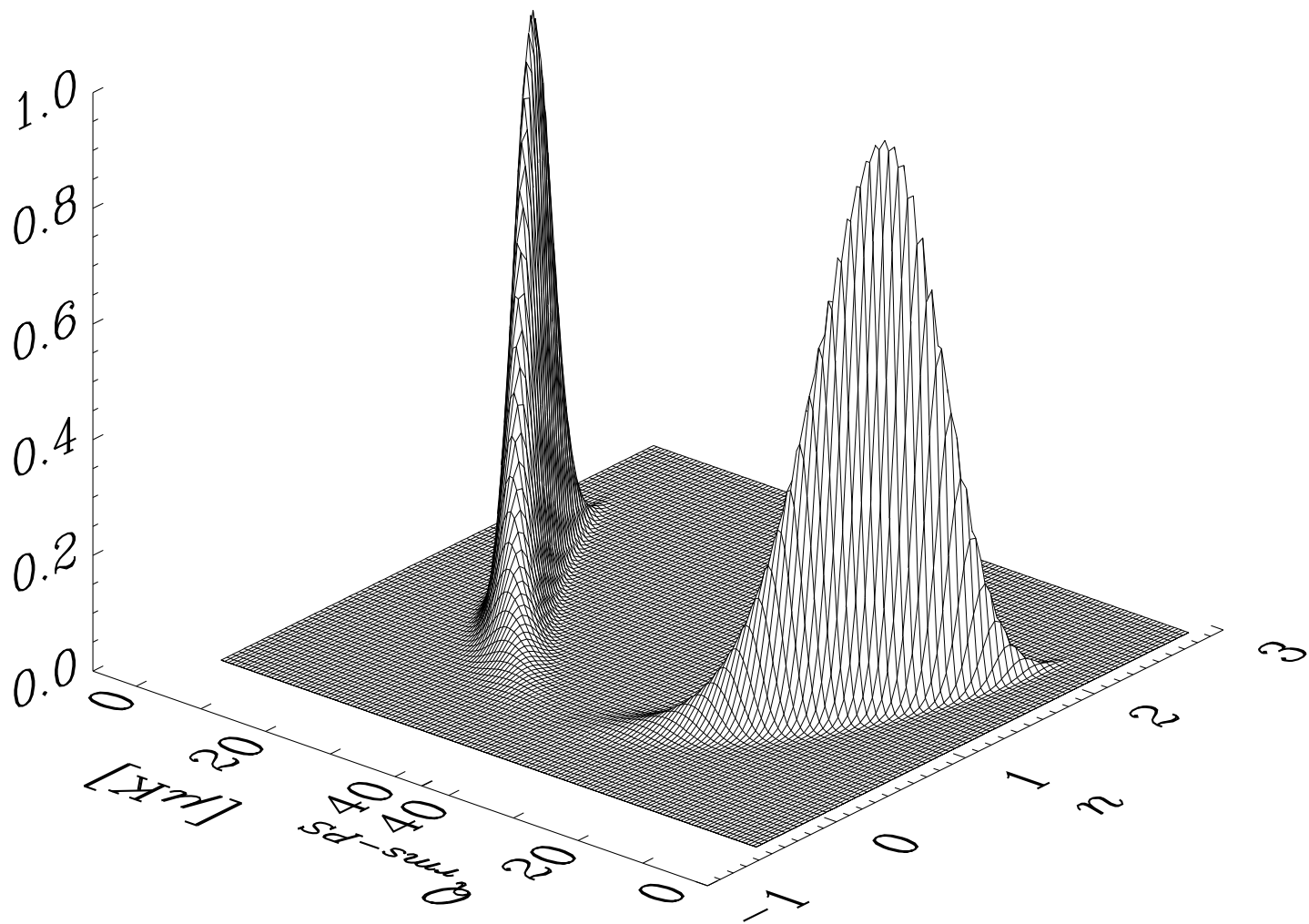

Fig. 1.— Two views of the likelihood function, $P(n, Q_{rms-PS})$, derived in a simultaneous analysis of the 53 and 90 GHz $COBE$-DMR two year data including harmonic amplitudes from the range $\ell \in [2, 30]$.

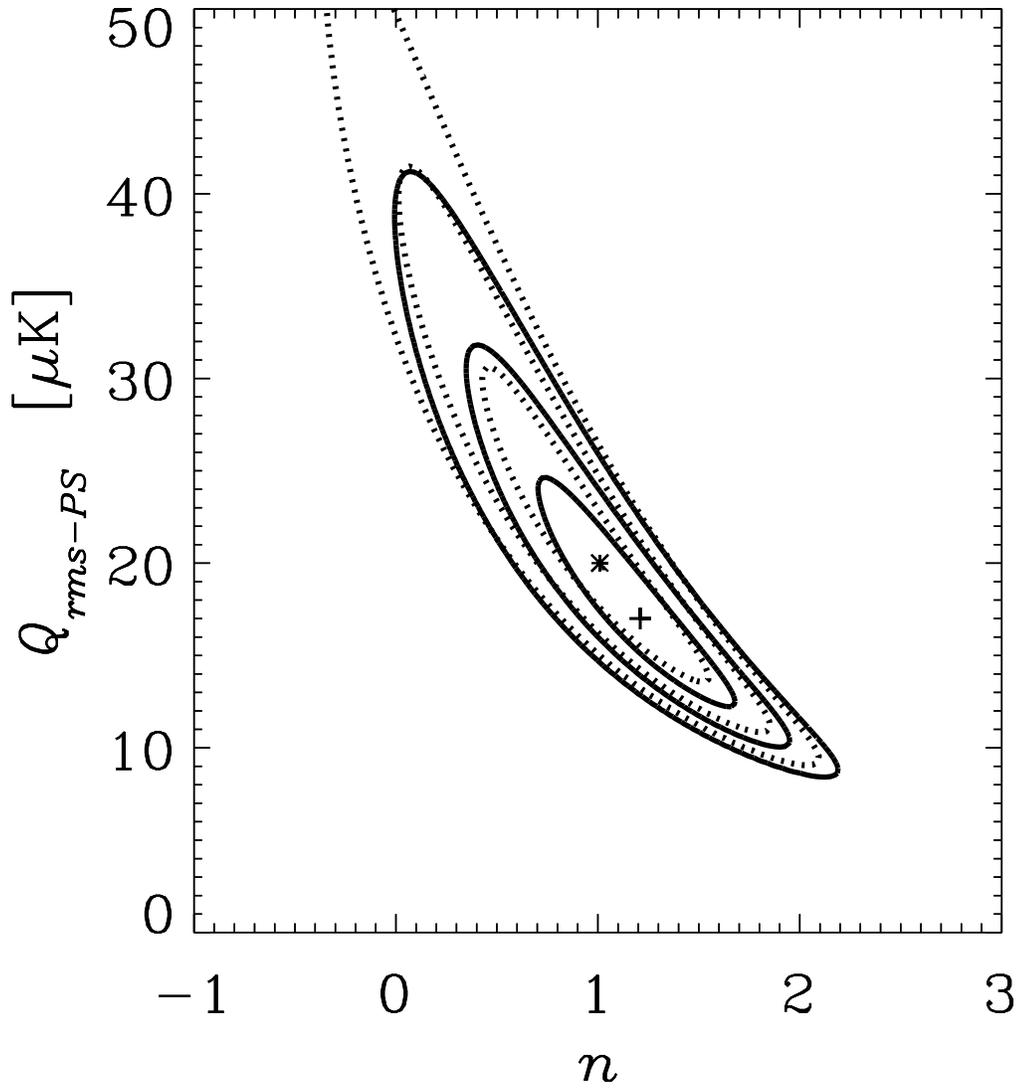

Fig. 2.— Contour plots of the likelihood function $P(n, Q_{rms-PS})$. 68%, 95% and 99.7% levels and the maximum values are shown. Solid lines and '+' represent the results of the analysis including the quadrupole. Dotted lines and '*' show the no quadrupole case.

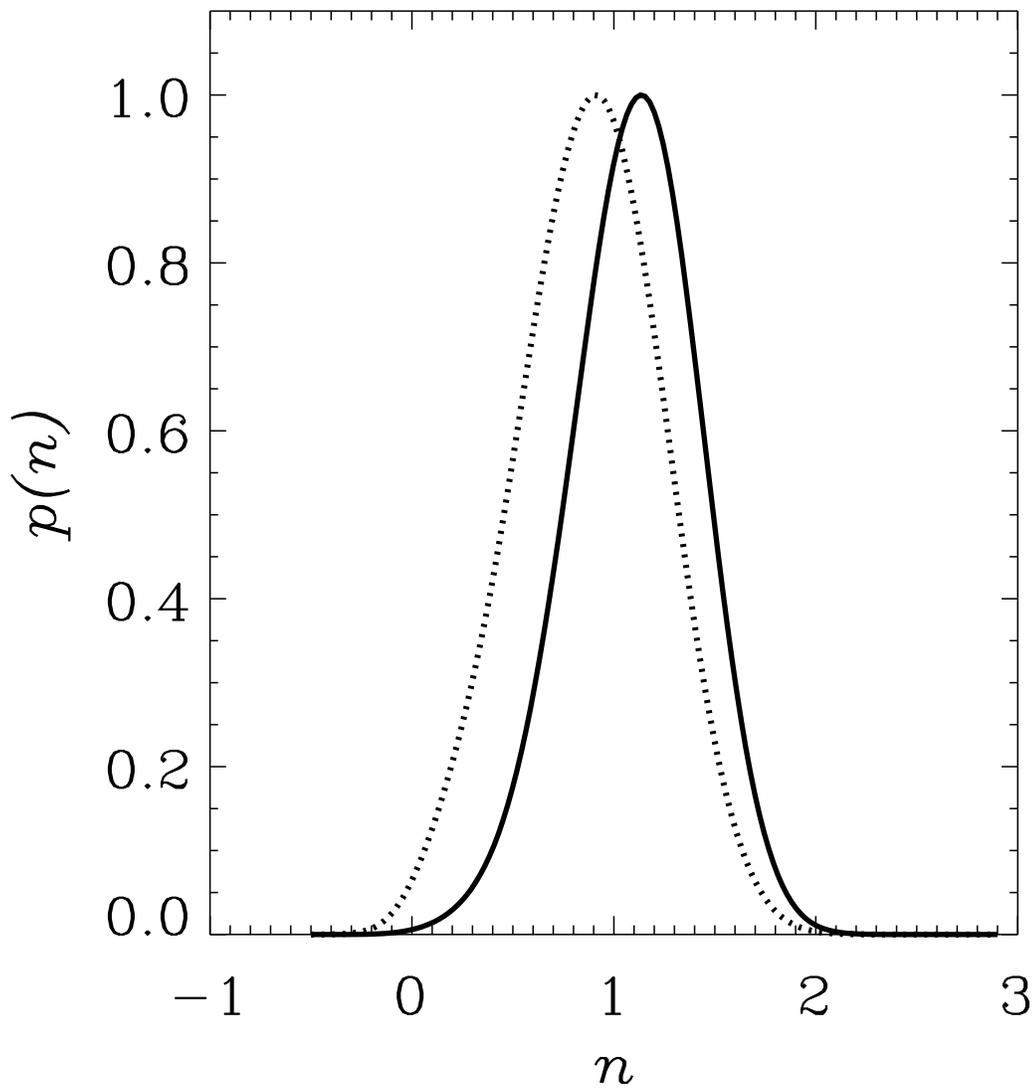

Fig. 3.— Marginal likelihood densities, $p(n) \propto \int dQ_{rms-PS}\, P(n, Q_{rms-PS})$, for the spectral index $n$ (arbitrary normalization). The solid curve includes the quadrupole in the analysis, the dotted curve excludes the quadrupole.

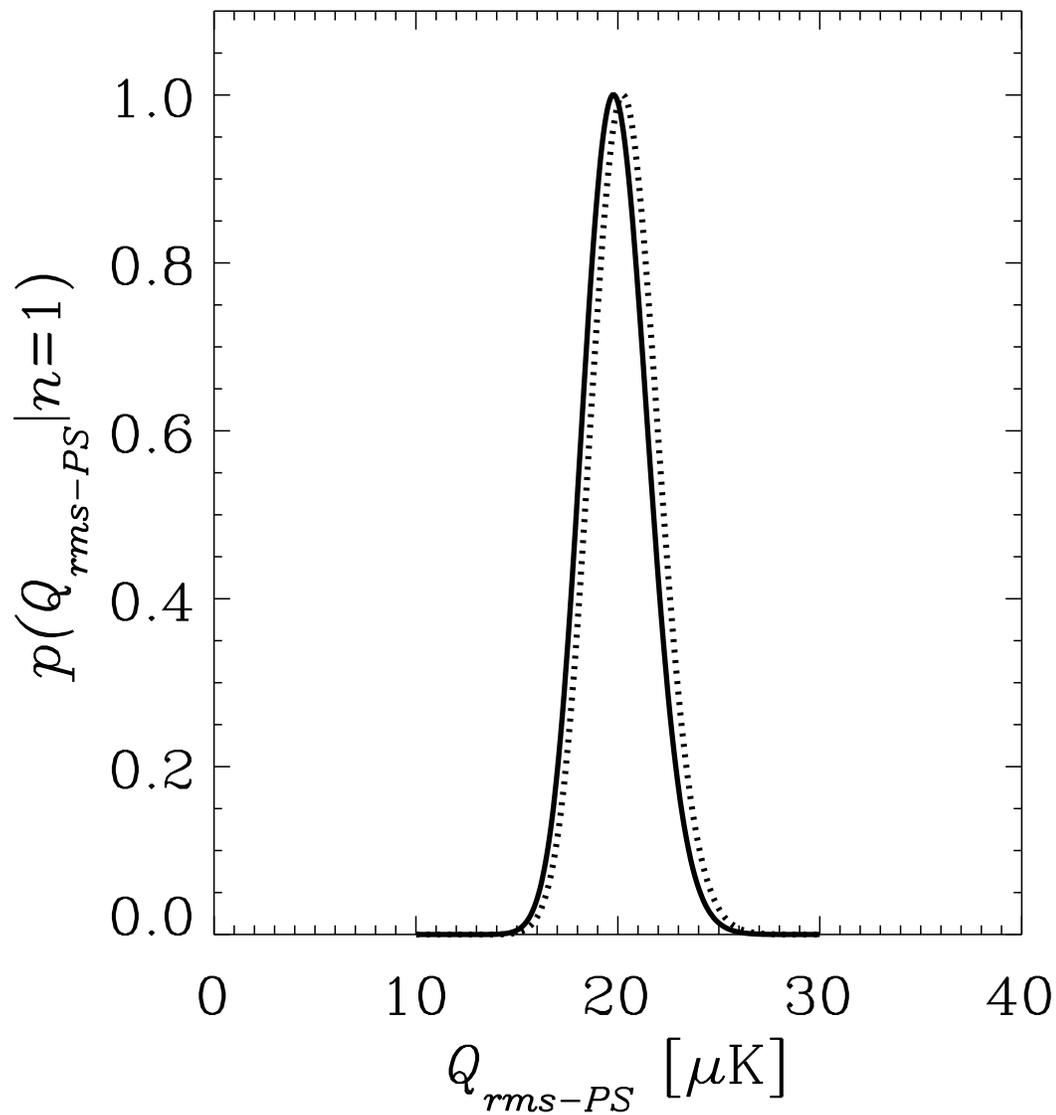

Fig. 4.— The $n = 1$-conditional likelihood density (arbitrary normalization) for $Q_{rms-PS}$. The solid curve includes the quadrupole, the dotted curve is the no quadrupole case.